\documentclass[a4paper,11pt]{article}
\usepackage[T1]{fontenc}
\usepackage[english]{babel}
\usepackage[intlimits]{amsmath}
\usepackage[ruled,vlined,linesnumbered]{algorithm2e}
\DontPrintSemicolon
\SetKwInput{KwIn}{Input}
\SetKwInput{KwOut}{Output}
\usepackage{amsthm}
\usepackage{amssymb}
\usepackage{eurosym}
\usepackage{float} 
\usepackage{xcolor}
\usepackage[sfdefault]{carlito}
\usepackage{newpxmath}
\usepackage{url}
\usepackage[pdftex]{graphicx}
\usepackage[table]{xcolor}
\usepackage{enumitem}
\setlist{nosep}
\usepackage{fontawesome}

\usepackage[colorlinks=true,linkcolor=blue,citecolor=blue,urlcolor=blue,pdfborderstyle={/S/U/W 1}]{hyperref}
\usepackage[backend=bibtex,style=authoryear,maxbibnames=30,maxcitenames=2,doi=true,uniquename=false,uniquelist=false]{biblatex}
\DeclareFieldFormat{doi}{\ifhyperref{\href{https://doi.org/#1}{\textcolor{blue}{doi: #1}}}{doi: #1}}
\DeclareFieldFormat{citeblue}{\DeclareFieldAlias{bibhyperref}{noformat}\bibhyperref{\textcolor{blue}{#1}}}
\renewcommand*{\multicitedelim}{\addsemicolon\space}
\DeclareCiteCommand{\parencite}{\usebibmacro{prenote}\bibopenparen}
  {\printtext[citeblue]{\printnames{labelname},~\printfield{year}\usebibmacro{postnote}}}
  {\multicitedelim}{\bibcloseparen} 
\DeclareCiteCommand{\textcite}{\usebibmacro{prenote}}
  {\printtext[citeblue]{\printnames{labelname}~(\printfield{year}\usebibmacro{postnote})}}
  {\multicitedelim}{}
\usepackage{setspace}
\usepackage[left=15mm,top=20mm,bottom=20mm,right=15mm]{geometry}
\usepackage{titlesec}
\titleformat*{\section}{\large\bfseries} 
\titleformat*{\subsection}{\bfseries}
\titlespacing{\section}{0cm}{6pt}{0pt}
\titlespacing{\subsection}{0cm}{6pt}{0pt}
\raggedright
\usepackage{fancyhdr}
\usepackage{lineno}
\usepackage[none]{hyphenat}
\usepackage{changepage}
\usepackage{orcidlink}

\begin{document}

\textbf{\huge{Three-dimensional inversion of gravity data using implicit neural representations and scientific machine learning}}

\vspace{0.5cm}
\large \textbf{Pankaj K Mishra} \orcidlink{0000-0003-4907-4724}$^{1\ast}$, \textbf{Sanni Laaksonen}$^{1}$, \textbf{Jochen Kamm}$^{1}$, \textbf{Anand Singh}$^{2}$

\vspace{0.2cm} 
\small
$^{1}$Geological Survey of Finland, Vuorimiehentie 5, Espoo, Finland\\
$^{2}$Indian Institute of Technology Bombay, Mumbai, India\\
$^{\ast}$Corresponding author: \href{mailto:pankaj.mishra@gtk.fi}{pankaj.mishra@gtk.fi}


\noindent\rule{\textwidth}{0.2pt}
\begin{adjustwidth}{4cm}{0cm} 
    \noindent\textbf{Abstract}\\
Inversion of gravity data is an important method for investigating subsurface density variations relevant to mineral exploration, geothermal assessment, carbon storage, natural hydrogen, groundwater resources, and tectonic evolution. Here we present a scientific machine-learning approach for three-dimensional gravity inversion that represents subsurface density as a continuous field using an implicit neural representation (INR). The method trains a deep neural network directly through a physics-based forward-model loss, mapping spatial coordinates to a continuous density field without predefined meshes or discretisation. Spatial encoding enhances the network’s capacity to capture sharp contrasts and short-wavelength features that conventional coordinate-based networks tend to oversmooth due to spectral bias. We demonstrate the approach on synthetic examples including smooth models, representing realistic geological complexity, and a dipping block model to assess recovery of structures at different depths. The INR framework reconstructs detailed structure and geologically plausible boundaries without explicit regularisation or depth weighting, while reducing the number of inversion parameters as the problem size grows bigger. These results highlight the potential of implicit representations to enable scalable, flexible, and interpretable large-scale geophysical inversion. This framework could generalise to other geophysical methods and for joint/multiphysics inversion.  \vspace{0.25cm}
    \\
    \textbf{Keywords}: Neural fields; gravity; physics-based deep learning; scientific machine learning, inversion 
\end{adjustwidth}

\noindent\rule{\textwidth}{0.2pt}


\section*{Introduction}
Inversion of gravity data to estimate subsurface density distributions is a well-known ill-posed problem in geophysics, as the measured gravity anomaly can be explained by infinitely many density configurations \parencite{parker1975theory}. Additional constraints or regularization are therefore required to obtain geologically reasonable solutions. Traditional 3D gravity inversion schemes discretize the subsurface into rectangular voxels (prisms) and solve for the density of each cell by minimizing a misfit to observed data, subject to stabilizing regularization terms. The seminal approach of \textcite{li19983} introduced a Tikhonov regularization with depth weighting to counteract the decay of sensitivity with depth, while later variants incorporated constraints promoting smoothness or compactness in the recovered model \parencite{last1983compact}.
In voxel-based parameterisation, rectangular cells seldom align with geologically plausible interfaces \parencite{danaei20223d}. Near the surface, this geometric mismatch can produce grid-aligned artefacts in the residuals, reflecting the staircase approximation of curved boundaries; at depth, it can suggest structure that the data do not truly support unless strong regularisation is imposed. The fixed grid also distributes degrees of freedom unevenly, that is, too few where the data are informative and too many where they are weak, so depth weighting or heavy regularisation is often introduced, which can suppress deeper features \parencite{oldenburg1991inversion,last1983compact}. More flexible and balanced representations such as unstructured mesh \parencite{danaei20223d} or octree-mesh \parencite{davis2013} place model freedom where the data support it and reduce reliance on depth weighting. 

With recent advances in machine learning and artificial intelligence, it has become increasingly intriguing to explore how traditional geophysical inversion can benefit from these developments without compromising physical rigour or interpretability. Deep learning methods have shown remarkable potential in many fields, and their application to geophysical inversion has accelerated in recent years. Data-driven approaches learn a direct mapping from gravity data to voxel models using large libraries of synthetic examples. \textcite{huang2021deep} used a fully convolutional network to predict 3D density anomalies; related work explores CNNs \parencite{cai2025effective}, U-Nets \parencite{yu2021three,wu2023improved,10034785}, encoder–decoder designs \parencite{yang20213,yang2022deep,9965416}, and decomposition networks such as DecNet \parencite{zhang2022decnet}. These methods typically require extensive training data, can struggle to generalise beyond the training distribution and usually output voxel models. As we move toward large-scale digital representations of the subsurface, developing  approaches where traditional physics is used to supervise the training of the neural network, is another alternative approach to mix domain knowledge and machine learning for solving geophysical inverse problems \parencite{Mishra2025}.  

In this paper we propose a 3D inversion framework using an implicit neural representation (also called a coordinate-based network or neural field), which encodes a continuous mapping from spatial coordinates to property (density) values \parencite{sitzmann2020implicit}. INRs have shown encouraging results in computer graphics for continuous images and 3D scenes \parencite{essakine2025}, and have recently appeared in geoscience applications, including 1D/2D seismic inversion \parencite{sun2023implicit,romero2025bayesian}, potential-field data processing \parencite{smith2025implicit}, geological modelling \parencite{hillier2023geoinr}, gravity-field modelling of irregular bodies in planetary science \parencite{izzo2022geodesy,schuhmacher2023investigation}, synthetic 2D inversions of DC and seismic data \parencite{xu2025towards}. A recent independent study has applied similar approach to 3D gravity inversion \parencite{li2025implicit}. In our setting, the INR serves as the model parametrisation: its weights define the 3D density field and are estimated by minimising the misfit between observed and forward-modelled gravity responses. With suitable spatial encodings, the representation can express sharp contrasts, while architectural bias provides an implicit regulariser that favours simple structures. 

Building on this representation, we perform \textit{physics-based INR inversion}, where the network is trained directly against the forward gravity operator rather than through pre-computed examples. Spatial coordinates are supplied to the network using positional encoding to address the spectral bias of standard multilayer perceptrons, which tend to learn low-frequency content more readily than high-frequency detail \parencite{rahaman2019spectral}. A multi-frequency sinusoidal encoding of the coordinates injects high-frequency basis functions that enable the network to represent sharp interfaces and small-scale heterogeneity, as demonstrated for images and 3D scenes by \textcite{tancik2020fourier}. In this context, spatial encoding provides the network a rich set of basis functions to combine, effectively mitigating spectral bias and enabling reconstruction of geologically realistic sharp density anomalies (e.g. edges of high-density bodies) that would otherwise be smeared out. We discuss two encoding strategies: positional encoding and multi-resolution hash encoding.  The network favours parsimonious and spatially coherent fields unless the data require more detail. Such implicit regularisation is also observed when using stochastic optimization for geophysical inversion with sparse parameterisation \parencite{mishra2025stochastic}.
 
In this study, we examine three main questions. First, can an INR, trained only with a data-misfit objective, act as an implicit regulariser that produces stable and geologically reasonable density fields? Second, how do the spatial encoding bandwidth and network size control the balance between smoothness and detail, and how does this relate to the resolution of the data? Third, does the compact neural parameterisation reduce the number of effective parameters compared to voxel models without losing model fidelity? We test these ideas using synthetic examples: Gaussian random fields to represent complex geological variability and a dipping block to assess the recovery of sharp contrasts. We then discuss where the approach performs well, limitations, and possible directions for future improvement. 


\section*{Methods}
\subsection*{Forward modelling of gravity data}
We consider the gravity‐inversion problem in which the observed data are gravity anomalies $g_{\text{obs}}(x,y)$ measured on a horizontal plane (e.g.\ at the surface) due to an unknown subsurface density contrast distribution $\rho(x,y,z)$. The forward relationship follows directly from Newton’s law of gravitation integrated over the volume. For practical computation, we discretize the model domain into $N_x \times N_y \times N_z$ rectangular cells (prisms) of constant density. The gravity contribution of each cell at an observation point can be evaluated analytically with the rectangular‐prism formula; here we adopt the formulation of \parencite{Nagy1966,nagy2000gravitational} for the vertical component $g_z$, which is implemented efficiently in our code. In matrix form, the forward model reads
\begin{equation}
\mathbf{d} = \mathbf{G}\mathbf{m},
\end{equation}

where $\mathbf{d}$ is an $N_{\text{obs}}$-element  column vector of gravity anomaly at the observation points, $\mathbf{m}$  is an $(N_{\text{cell}}$-element column vector of cell densities, and $\mathbf{G}$ is the $N_{\text{obs}} \times N_{\text{cell}}$ sensitivity matrix.  Each element $G_{ij}$ represents the gravity effect  at observation~$i$ due to a unit‐density perturbation in cell~$j$. We compute $\mathbf{G}$ once at the start by summing the contributions from the eight prism corners (via the usual arctan/logarithmic terms). In our experiments the observation points lie on a regular $40\times 40$ surface grid, aligned (at the centres of the surface cells) with the model grid for simplicity, though the formulation supports arbitrary survey layouts. Once $\mathbf{G}$ is known, forward modeling reduces to a matrix–vector multiplication.

\subsection*{Mathematical formulation of the inverse problem}
The gravity inversion is formulated as an optimisation problem in the parameter space of the neural network. Let $\rho_\theta(\mathbf{x})$ denote the density contrast field represented by an implicit neural representation with parameters $\theta$, evaluated at spatial location $\mathbf{x}$. The forward operator $\mathbf{G}$ maps the discretised density field (evaluated at cell centres) to predicted gravity data.
\\
The inverse problem is defined as the minimisation of a weighted data-misfit functional:
\begin{equation}
\theta^\star = \arg\min_{\theta} \; 
\left\| \mathbf{W}_d \left( \mathbf{G}\,\mathbf{m}(\theta) - \mathbf{d}_{\mathrm{obs}} \right) \right\|_2^2,
\end{equation}
where $\mathbf{m}(\theta)$ is the vector of densities obtained by evaluating $\rho_\theta$ at all cell centres, $\mathbf{d}_{\mathrm{obs}}$ is the observed gravity data, and $\mathbf{W}_d$ is a data-weighting matrix, typically chosen as $\mathbf{W}_d = \sigma^{-1}\mathbf{I}$ under the assumption of uncorrelated noise with standard deviation $\sigma$.
\\
The optimisation variables are the neural-network parameters $\theta$, which define the continuous density field. No explicit model objective (regularisation term) is included. Instead, stability is controlled implicitly through the network parametrisation (architecture and encoding), bounded output activation, and early stopping based on a discrepancy-principle criterion. This formulation contrasts with classical deterministic inversion, where explicit regularisation terms are introduced to control the null space and stabilise the solution.

For a realistic test, we generate a synthetic \textit{true} density model $\rho_{\text{true}}(x,y,z)$ by sampling a Gaussian random field (GRF) on the 3D grid \parencite{liu2019advances}.  After sampling, the field is linearly rescaled to the range $1.6\text{–}3.5\;\text{g cm}^{-3}$, representative of common crustal contrasts (e.g.\ a $2.0\;\text{g cm}^{-3}$ background with embedded higher‐ and lower‐density bodies). From this $\rho_{\text{true}}$ we compute noise-free gravity data
$\mathbf{d}_{\text{true}}=\mathbf{G}\,\mathbf{m}_{\text{true}}$ and then add zero-mean Gaussian noise,
\begin{equation}
\mathbf{d}_{\mathrm{obs}}=\mathbf{d}_{\mathrm{true}}+\mathbf{n},\qquad
\mathbf{n}\sim\mathcal{N}\!\left(0,\ \sigma_n^2\mathbf{I}\right),
\end{equation}

with $\sigma_n=\alpha\,s_{\text{true}}$, where $\alpha$ is a prescribed noise level (0.01 in our tests), $\mu_{\mathrm{obs}}=\operatorname{mean}(\mathbf{d}_{\mathrm{obs}})$  and $s_{\text{true}}=\mathrm{std}(\mathbf{d}_{\text{true}})$.
For training stability, we standardise the data by subtracting the mean and dividing by the standard deviation of the observed data,
\begin{equation}
\tilde{\mathbf{d}}_{\mathrm{obs}}=\frac{\mathbf{d}_{\mathrm{obs}}-\mu_{\mathrm{obs}}}{s_{\mathrm{obs}}},
\end{equation}
and we apply the same affine transform to the forward predictions before computing the loss.

\subsection*{Implicit Neural Representation and Spatial Encodings}
\begin{figure}[H]
    \centering
    \includegraphics[width=\textwidth]{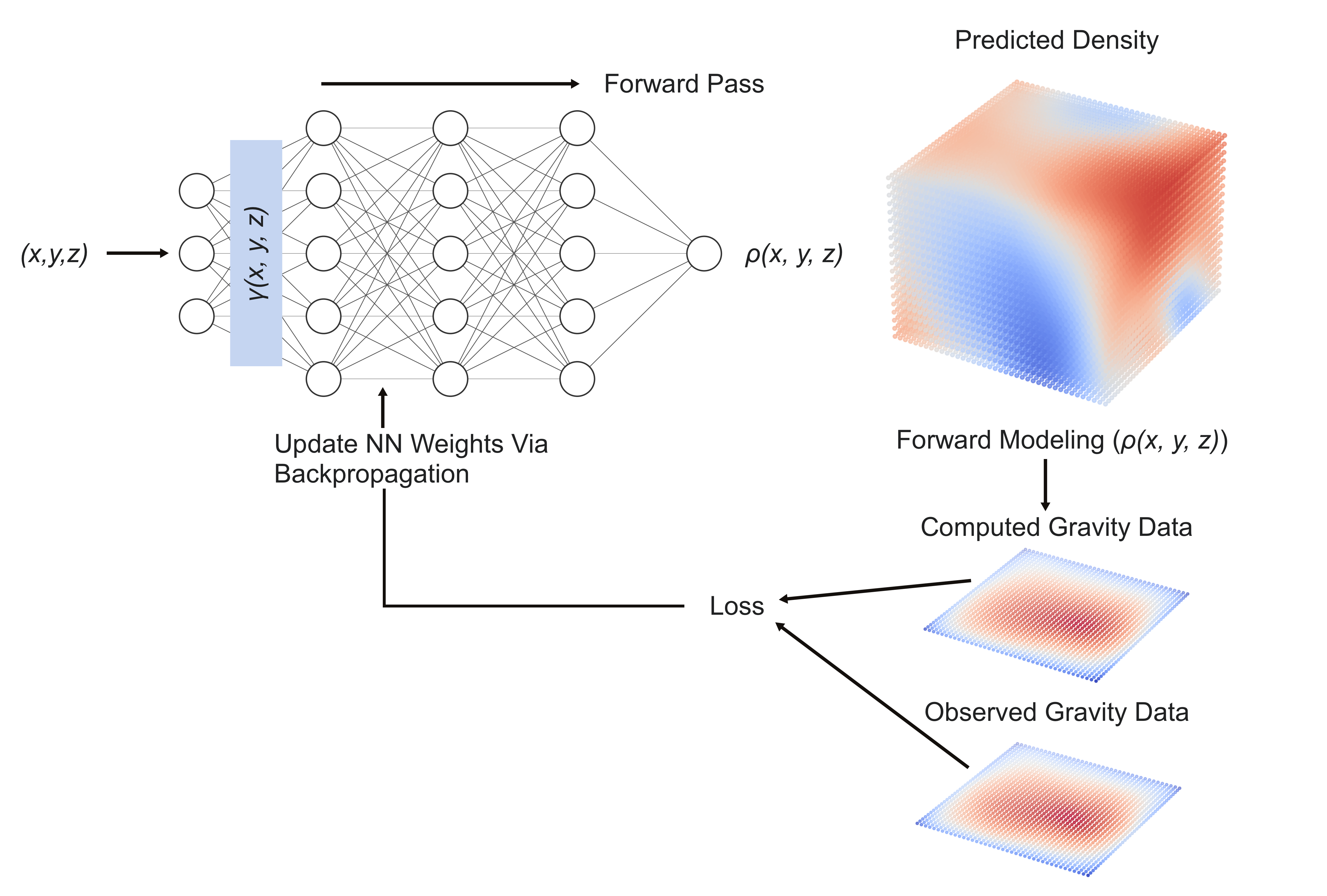}
    \caption{Physics-based machine learning inversion framework}
    \label{fig:Network}
\end{figure}

We represent density as a continuous implicit function $\rho_\theta(\mathbf{x})$ with trainable parameters $\theta$, implemented as a fully connected multilayer perceptron (MLP) (Figure~\ref{fig:Network}). The network input is the spatial coordinate $\mathbf{x}=(x,y,z)$ after standardising each component to be dimensionless by subtracting the dataset mean and dividing by its standard deviation. Throughout this study, the primary encoding is a fixed sinusoidal positional encoding \parencite{tancik2020fourier,mildenhall2021nerf}, chosen because it is simple, transparent, and easy to interpret in terms of frequency content. Since learnable multi-scale encodings have become increasingly common in recent INR studies, we also evaluate a multi-resolution hash encoding \parencite{muller2022instant} as an alternative benchmark in the revised experiments. In both cases, the encoding maps coordinates to a feature vector that is then passed to the same MLP backbone. 
\begin{enumerate}
    \item For the sinusoidal positional encoding, each scalar coordinate $u\in\{x,y,z\}$ is mapped to
\begin{equation}
\boldsymbol{\gamma}(u) 
=
\big[u,\ \cos(\beta\,2^{0}u),\ \sin(\beta\,2^{0}u),\ \ldots,\ \cos(\beta\,2^{n-1}u),\ \sin(\beta\,2^{n-1}u)\big],
\end{equation}
where $n$ is the number of dyadic frequency bands and $\beta$ is a bandwidth parameter. The full input to the MLP is $\boldsymbol{\gamma}(x)\,\|\,\boldsymbol{\gamma}(y)\,\|\,\boldsymbol{\gamma}(z)$, giving $3(1+2n)$ features per point. This encoding is fixed during training and enriches the coordinate input with basis functions spanning multiple spatial scales, which helps counter the spectral bias of standard coordinate MLPs.
    \item As an alternative benchmark, we also consider a multi-resolution hash encoding. In this representation, the coordinate $\mathbf{x}$ is mapped to a hierarchy of rectilinear grids with progressively finer spatial resolution. Each grid level stores learnable feature vectors at its corner nodes in a fixed-size hash table. For a query point, features are obtained at each level by trilinear interpolation of the eight neighbouring corner embeddings, and the interpolated features from all levels are concatenated and supplied to the MLP.
\end{enumerate}
Other coordinate encodings are also possible, however, we focus here on two approaches above because they span two widely used and conceptually distinct choices: a fixed frequency-based mapping and a learnable multiresolution mapping. We use LeakyReLU \parencite{xu2020reluplex} activations (slope $0.01$) and Adam optimisation \parencite{kingma2014adam}  throughout this paper. The output layer uses a bounded activation mapped to prescribed density bounds to keep predictions in a physically plausible range and to aid optimization. For absolute-density models, a sigmoid output is scaled to the specified density interval, whereas for density-contrast models a tanh output is scaled to the prescribed maximum contrast, yielding bounded predictions with symmetric support about zero.

In the inversion, the forward calculation uses rectangular prisms, and the loss is evaluated only at cell centres. We therefore treat $\rho_\theta$ evaluated at cell centres as the cell densities, so the effective model seen by the physics is a block model at the chosen forward discretisation. Values of $\rho_\theta$ at off-grid locations are not constrained by the loss and should be interpreted as an interpolation consistent with the chosen spatial encoding, useful for visualisation rather than as evidence of additional resolved structure. As the forward grid is refined and the prism response converges, the method approaches practically continuous behaviour. At coarser grids, genuinely continuous behaviour would require either a much finer forward discretisation or a forward operator that permits within-cell variability, for example through polynomial or finite-element formulations. More generally, spatial encoding expands the coordinate input into a richer feature representation, which can improve the recovery of sharper contrasts when supported by the data, while the network architecture itself favours parsimonious and spatially coherent fields. In practice, the effective smoothness of the recovered model depends on both the encoding and the network capacity: a high-capacity network with a rich spatial encoding can reproduce more oscillatory structure, whereas a smaller network or a smoother encoding yields a smoother approximation and thus acts as an implicit regulariser.

We clarify that the framework used here is \emph{physics-based/physics-guided/physics-aware machine learning}, within the broader area of scientific machine learning. The network is trained by minimising a \emph{physics-based forward-model loss}: the differentiable gravity operator maps a candidate density field to predicted data, and the data misfit drives learning. We do \emph{not} enforce governing PDE residuals, boundary conditions, or collocation losses as in PINNs. In short, this is a data-misfit–driven inversion with a neural \emph{representation} of the model, not a PINN solving a PDE. For a broader discussion of nomenclature and flavours of scientific machine learning, see \textcite{Faroughi2024}. Following is a high-level pseudocode describing how this inversion framework is implemented; we encourage readers to see the open-source python implementations provided with this manuscript to understand technical implementation details:

\begin{algorithm}[H]
\caption{Physics-based INR inversion with  spatial encoding}
\KwIn{
\textbf{Cell-centre coordinates:} $\mathbf{x}_j=(x_j,y_j,z_j)$, $j=1,\dots,N_{\mathrm{cell}}$;\\
\textbf{Observation coordinates:} $\mathbf{x}^{\mathrm{obs}}_i$, $i=1,\dots,N_{\mathrm{obs}}$;\\
\textbf{Observed gravity:} $g_{z}^{\mathrm{obs}}$ (or synthetic $g_{z}^{\mathrm{obs}}$ generated from a reference model plus noise);\\
\textbf{Forward discretisation:} prism dimensions $(d_x,d_y,d_z)$ and sensitivity matrix $\mathbf{G}$;\\
\textbf{Encoding choice:} $\mathcal{E}\in\{\text{positional},\text{hash}\}$ with hyperparameters $\eta$;\\
\textbf{Network and training:} hidden width, depth, bound $\rho_{\max}$, learning rate, epochs $T$, data weight $W_d$, loss weight $\gamma$.
}
\KwOut{
Trained parameters $\theta^\star$; recovered density contrast $\rho_{\theta^\star}$ at cell centres; predicted gravity $g_z^{\mathrm{pred}}$.
}

\BlankLine
\textbf{\faCode\ Prepare geometry}\\
\quad Build cell-centre grid and observation coordinates.\\
\quad Standardise coordinates componentwise:
$u \leftarrow (u-\mu_u)/\sigma_u$, $\; u\in\{x,y,z\}$.

\BlankLine
\textbf{\faCode\ Assemble forward operator}\\
\quad Compute the prism-based sensitivity matrix $\mathbf{G}$ once from the rectangular-prism formula.

\BlankLine
\textbf{\faCode\ Prepare data}\\
\quad If using a synthetic test, define a reference density model $\rho_{\mathrm{true}}$, compute
$g_z^{\mathrm{true}}=\mathbf{G}\rho_{\mathrm{true}}$, add Gaussian noise, and set
$g_z^{\mathrm{obs}}=g_z^{\mathrm{true}}+n$.\\
\quad Estimate the noise scale $\sigma$ and set the scalar data weight $W_d=1/\sigma$.

\BlankLine
\textbf{\faCode\ Define encoded INR model}\\
\quad Choose spatial encoding $\mathcal{E}(\mathbf{x};\eta)$
(e.g.\ sinusoidal positional encoding or multi-resolution hash encoding).\\
\quad Define an MLP with trainable parameters $\theta$.\\
\quad For each cell centre, predict density contrast as
$\rho_\theta(\mathbf{x})=\rho_{\max}\tanh(\mathrm{MLP}(\mathcal{E}(\mathbf{x};\eta)))$.

\BlankLine
\textbf{\faCode\ Optimisation loop}\\
\For{$t=1$ \KwTo $T$}{
  \quad Evaluate $\rho_\theta$ at all cell centres.\\
  \quad Forward model:
  $g_z^{\mathrm{pred}}=\mathbf{G}\rho_\theta$.\\
  \quad Residual:
  $r=g_z^{\mathrm{pred}}-g_z^{\mathrm{obs}}$.\\
  \quad Loss:
  $\mathcal{L}= \frac{1}{N_{\mathrm{obs}}}\sum_{i=1}^{N_{\mathrm{obs}}}\left[W_d\left(g_{z,i}^{\mathrm{pred}}-g_{z,i}^{\mathrm{obs}}\right)\right]^2$.\\
  \quad Backpropagate $\mathcal{L}$ and update $\theta$ with Adam.}

\BlankLine
\textbf{\faCode\ Return and evaluate}\\
\quad Set $\theta^\star\leftarrow\theta$ and return $\rho_{\theta^\star}$ and $g_z^{\mathrm{pred}}$.\\
\quad Compute summary metrics such as RMS density-contrast error and RMS data misfit.\\
\quad Use off-grid evaluations of $\rho_{\theta^\star}(\mathbf{x})$ only for visualisation.
\end{algorithm}


\section*{Results}
\subsection*{Spectral bias and spatial encoding comparison}
\begin{figure}[H]
    \centering
    \includegraphics[width=\linewidth]{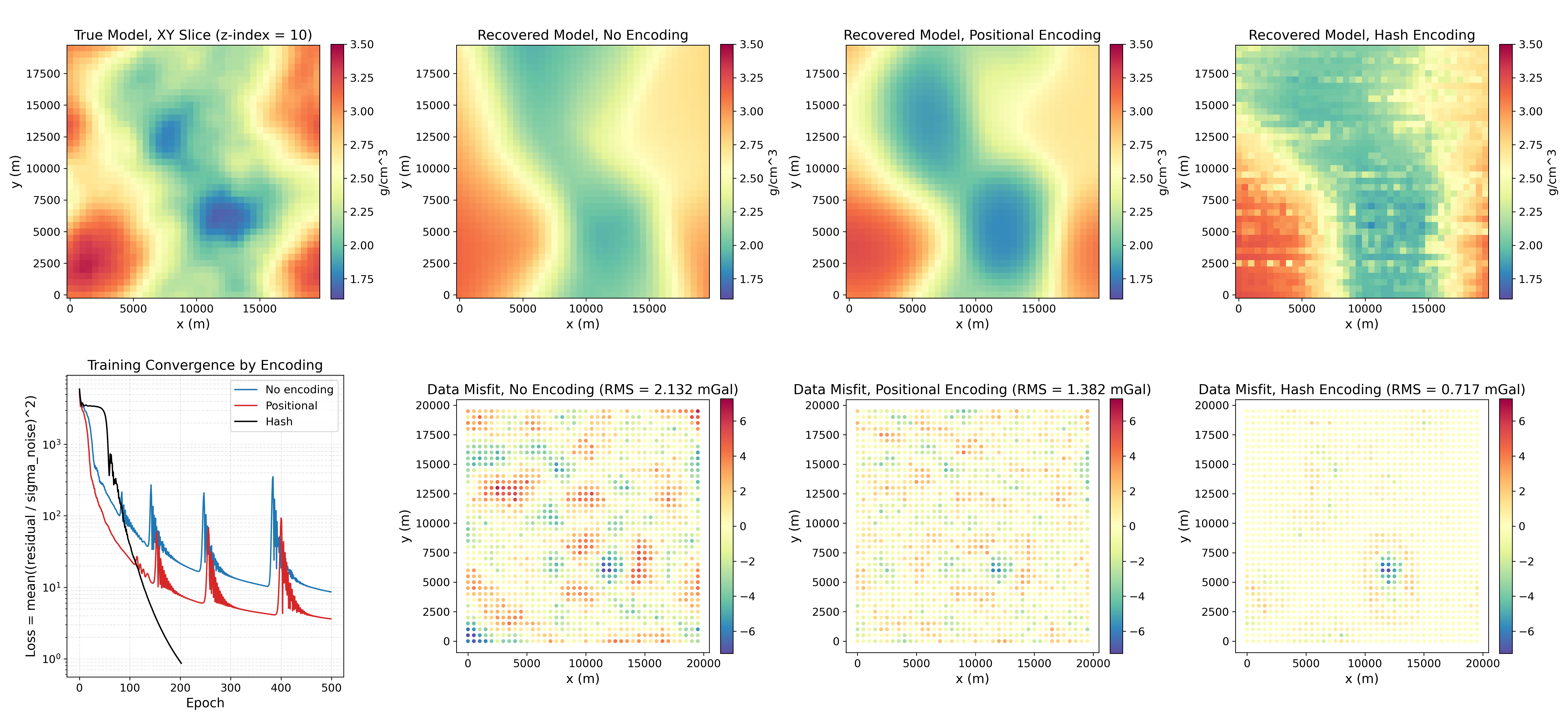}
    \caption{Comparison of inversion results obtained with no encoding, sinusoidal positional encoding, and multi-resolution hash encoding for a smooth Gaussian random field density model. Top row: true and recovered mid-depth density slices. Bottom row: training-loss histories and gravity-data residual maps.}
    \label{fig:EncodingComparisonSmooth}
\end{figure}

The purpose of this numerical test is to examine how the choice of spatial encoding affects representational bandwidth, convergence, and model-space behaviour in 3D INR gravity inversion. We generate a stationary Gaussian random field (GRF) on a $40\times40\times20$ grid with $d_x=d_y=d_z=500~\mathrm{m}$ (footprint $19.5\times19.5~\mathrm{km}$, depth $9.5~\mathrm{km}$), and linearly rescale it to the density range $[1.6,\,3.5]~\mathrm{g\,cm^{-3}}$. Vertical gravity $g_z$ is forward modelled at $40\times40$ surface stations using the rectangular-prism kernel, and Gaussian noise with standard deviation $\sigma_n=0.01\,\mathrm{std}(\mathbf{d}_{\mathrm{true}})$ is added. Three INR models are then trained under matched inversion settings: (i) a plain coordinate MLP with no encoding, (ii) the same MLP with inclusive sinusoidal positional encoding, and (iii) the same MLP with multi-resolution hash encoding. In all three cases, the MLP backbone is identical: four hidden layers of width $256$, LeakyReLU activations with slope $0.01$, and a sigmoid output mapped to $[1.6,\,3.5]~\mathrm{g\,cm^{-3}}$. The only difference between the three models is the coordinate representation supplied to the network. The positional encoding uses two dyadic frequency bands, while the hash encoding uses two resolution levels with two features per level, a hash-table size of $2^{17}$, base resolution $4$, and finest resolution $128$. All models are trained with Adam at learning rate $10^{-3}$ for up to $500$ epochs using the whitened data-misfit loss $\mathcal{L}=\mathrm{mean}[(r/\sigma_n)^2]$, with early stopping once the weighted misfit reaches the expected noise floor.

Figure~\ref{fig:EncodingComparisonSmooth} compares a mid-depth horizontal slice ($z$ index 10), the training-loss histories, and the gravity-data residual maps for the three encodings. The model without encoding shows the expected low-frequency bias of a plain coordinate MLP: it reproduces only broad long-wavelength structure, misses much of the intermediate-scale variability present in the true GRF, and leaves a strongly structured residual field with the largest RMS data misfit (2.132 mGal). Introducing sinusoidal positional encoding substantially improves the reconstruction. The recovered model follows the main morphology of the GRF more closely, captures the principal low-density and high-density regions in the correct locations, and reduces the RMS data misfit to 1.382 mGal. In this smooth-target example, the positional-encoding result also remains spatially coherent and visually consistent with the underlying GRF.

The multi-resolution hash encoding reaches the lowest final loss and the smallest RMS data misfit (0.717 mGal), and it approaches the noise floor in fewer epochs than the other two models. However, the recovered density slice exhibits visible banding, patchiness, and grid-aligned texture that are not present in the true smooth GRF. Thus, although the hash encoding explains the data best, it does not produce the most plausible recovered model for this particular target. The positional encoding gives a smoother and more coherent reconstruction, whereas the hash encoding behaves as a more flexible local representation that fits the data more aggressively.
\begin{figure}[H]
    \centering
    \includegraphics[width=\linewidth]{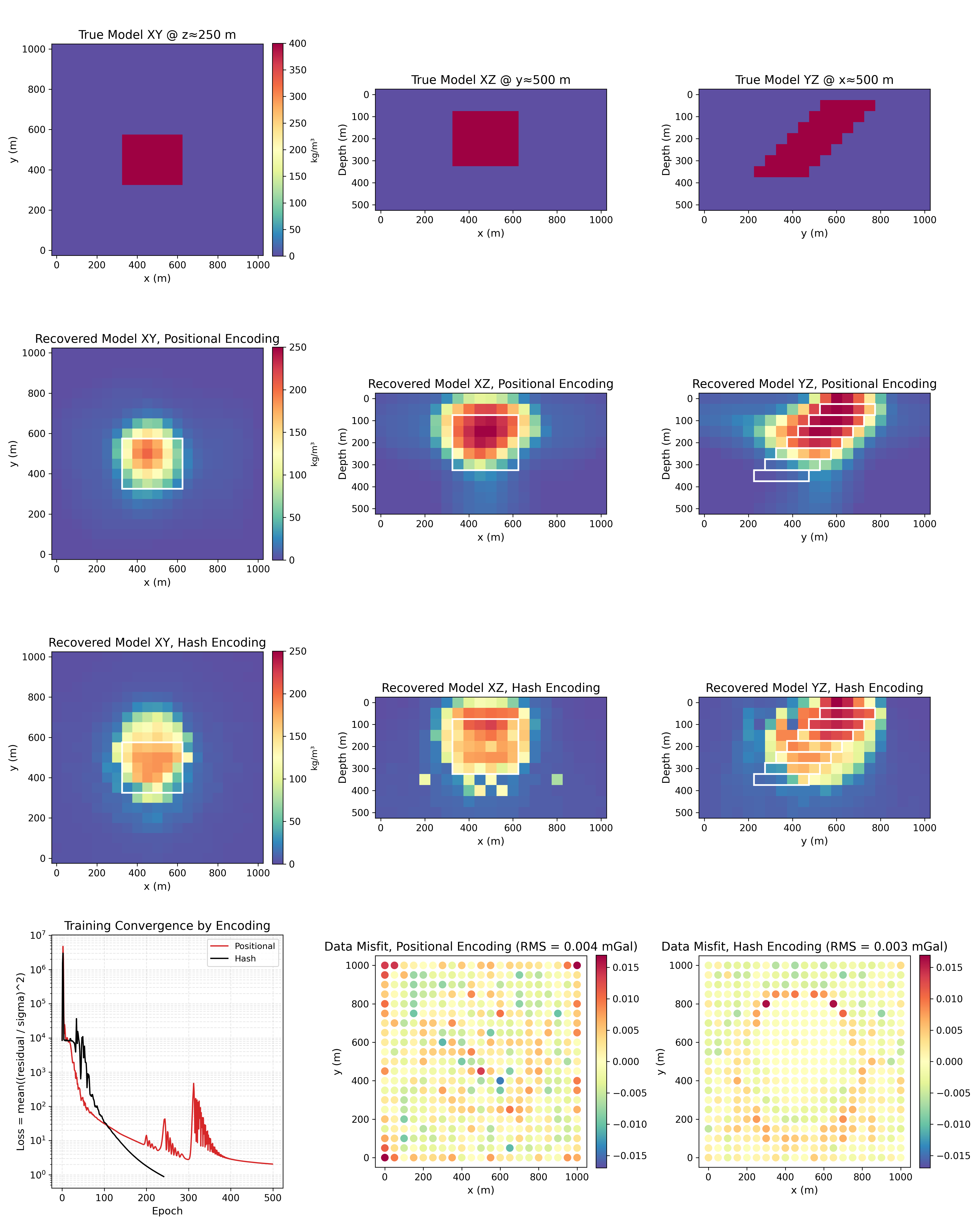}
    \caption{Comparison of inversion results for the dipping block-model test using no encoding, sinusoidal positional encoding, and multi-resolution hash encoding. Top row: true and recovered mid-depth density slices. Bottom row: training-loss histories and gravity-data residual maps.}
    \label{fig:EncodingComparisonBlock}
\end{figure}

As a complementary second test Figure~\ref{fig:EncodingComparisonBlock}, we repeated the encoding comparison on the dipping block model used in the blocky-structure experiment. This example differs from the smooth GRF benchmark in that the true density contrast is piecewise constant and contains abrupt geometric boundaries, while the gravity response remains band-limited by the forward kernel and survey geometry. The model volume is sampled on a $21\times21\times11$ grid with $d_x=d_y=d_z=50~\mathrm{m}$, and gravity $g_z$ is observed at $21\times21$ surface stations. Gaussian noise with standard deviation $\sigma_n=0.01\,\mathrm{std}(\mathbf{d}_{\mathrm{true}})$ is added. Two INR models are then compared under matched settings: the same four-layer width-$256$ MLP backbone, the same LeakyReLU activations, the same whitened data-misfit loss, the same Adam optimiser, and the same early-stopping criterion near the expected noise floor. To keep the comparison controlled, the same noise realisation and the same model-initialisation random state are used for both encodings. The sinusoidal positional encoding again uses two dyadic frequency bands, and the hash encoding uses two resolution levels with two features per level, a hash-table size of $2^{17}$, base resolution $4$, and finest resolution $128$.

In this blocky test, both encodings recover the main body location and the dipping trend in the $y$--$z$ section, and both fit the gravity data to near-noise level. The multi-resolution hash encoding reaches a slightly lower RMS data misfit (0.003 mGal versus 0.004 mGal) and converges somewhat faster, indicating that its greater local flexibility can be beneficial when the true model contains sharper interfaces. However, the recovered hash model still exhibits mild patchiness and isolated artefacts, particularly in the $x$--$z$ section, whereas the positional-encoding result remains smoother and cleaner in model space. Thus, even for a discontinuous target, the lower residual achieved by the hash encoding does not translate into an unambiguously better inversion model.

Taken together, the two encoding tests show that spatial encoding acts as part of the implicit prior of the inversion. Without encoding, the coordinate MLP is overly smooth and underfits the data. For the smooth GRF target, sinusoidal positional encoding provides the most favourable balance between detail recovery and spatial coherence, whereas multi-resolution hash encoding attains the lowest residual at the cost of visible texture and patchiness. For the blocky target, hash encoding becomes more competitive in data space and can recover somewhat sharper local structure, but it still introduces mild artefacts and does not clearly dominate positional encoding in model space. The preferred encoding in gravity inversion therefore depends on the character of the target model and should be judged in both data space and model space rather than by residual reduction alone. For the remaining experiments, we therefore use sinusoidal positional encoding with two frequency bands.

\subsection*{Network Capacity and Grid Refinement}

To better understand how implicit neural representations (INRs) behave in gravity inversion, we performed two complementary numerical experiments on the same block-model test problem. In the first experiment, we fixed the spatial discretisation and varied the INR network size. In the second, we fixed the INR architecture and varied the grid resolution. Taken together, these tests separate two effects that are coupled in conventional voxel inversion: representational capacity and discretisation.

\begin{figure}[H]
    \centering
    \includegraphics[width=\linewidth]{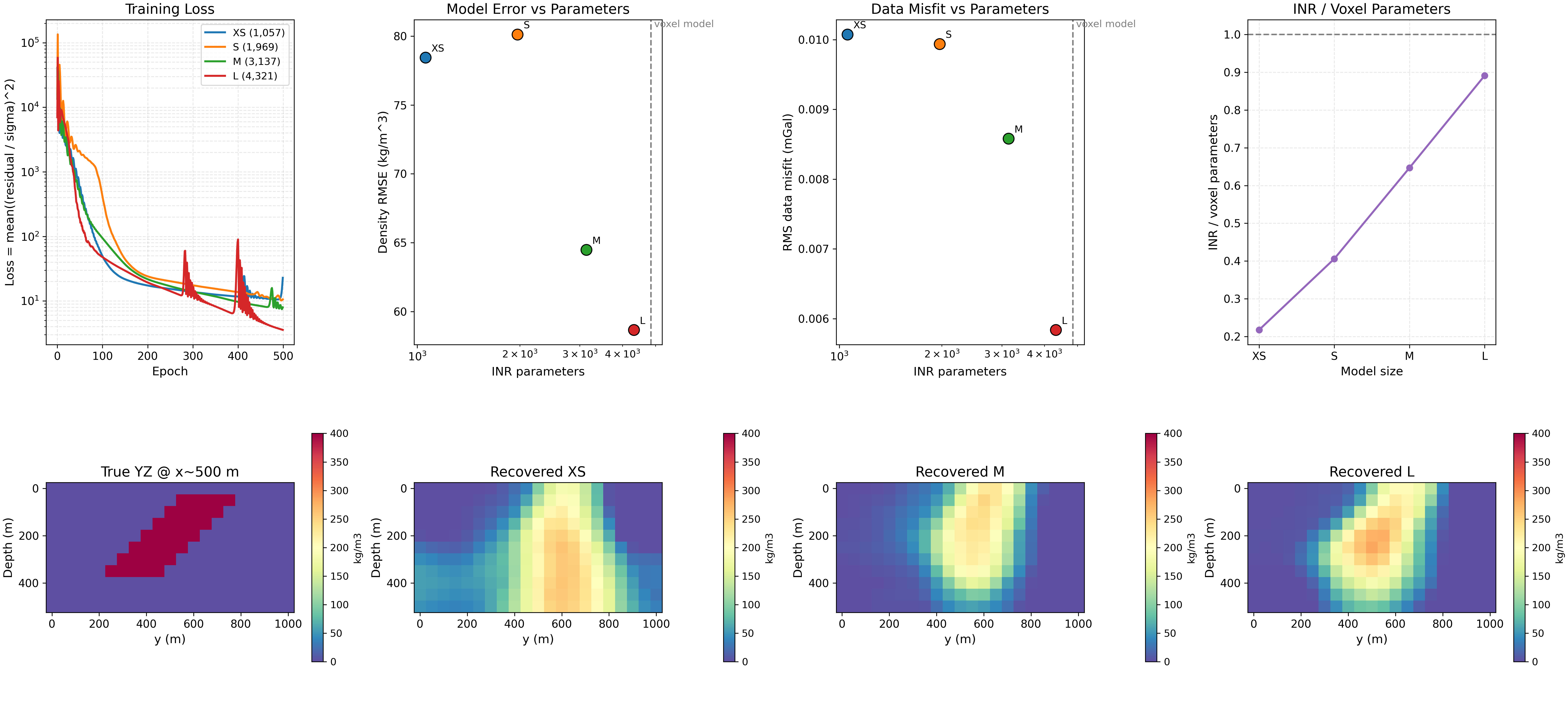}
    \caption{Effect of INR network size on 3D gravity inversion of the block model using positional encoding. Top row: training-loss histories, density-model RMSE versus INR parameter count, gravity-data RMSE versus INR parameter count, and the ratio of INR parameters to voxel-model parameters for the same discretisation. The dashed line marks the voxel-model parameter count. Bottom row: true $y$--$z$ cross-section of the block model together with recovered $y$--$z$ cross-sections for representative small, intermediate, and large INR architectures. As the network size increases, the recovered model becomes progressively more consistent with the dipping block geometry, while the model error and data misfit decrease, even though the largest INR still uses fewer parameters than the voxel model on this grid.}
    \label{fig:BlockNetworkSizeComparison}
\end{figure}

In the first test, the block-model discretisation is held fixed and only the size of the positional-encoding INR is varied. The synthetic target is a dipping block of density contrast $400~\mathrm{kg\,m^{-3}}$ embedded in a zero-contrast background over a $1000\times1000\times500~\mathrm{m}$ domain. Surface gravity $g_z$ is forward modelled at all surface grid locations and contaminated with Gaussian noise of standard deviation $\sigma_n = 0.01\,\mathrm{std}(\mathbf{d}_{\mathrm{true}})$. All inversions use the same positional encoding with two frequency bands, the same whitened data-misfit loss $\mathcal{L}=\mathrm{mean}[(r/\sigma_n)^2]$, Adam optimisation with learning rate $10^{-2}$, and the same early-stopping rule near the noise floor. The only quantity changed is the MLP architecture.

Figure~\ref{fig:BlockNetworkSizeComparison} shows that increasing INR size improves both the recovered model and the data fit, but the parameter growth is not tied to the number of grid cells in the same way as in a voxel formulation. The smallest networks underfit the block and recover only a broad, diffuse anomaly, whereas the intermediate and large networks progressively recover a sharper dipping structure and reduce both density-model RMSE and gravity-data RMS misfit. Importantly, the best-performing large INR still remains below the voxel-model parameter count on the same grid. This experiment therefore shows that INR inversion can gain representational power through network enlargement without requiring one-parameter-per-cell growth in the model space.

\begin{figure}[H]
    \centering
    \includegraphics[width=\linewidth]{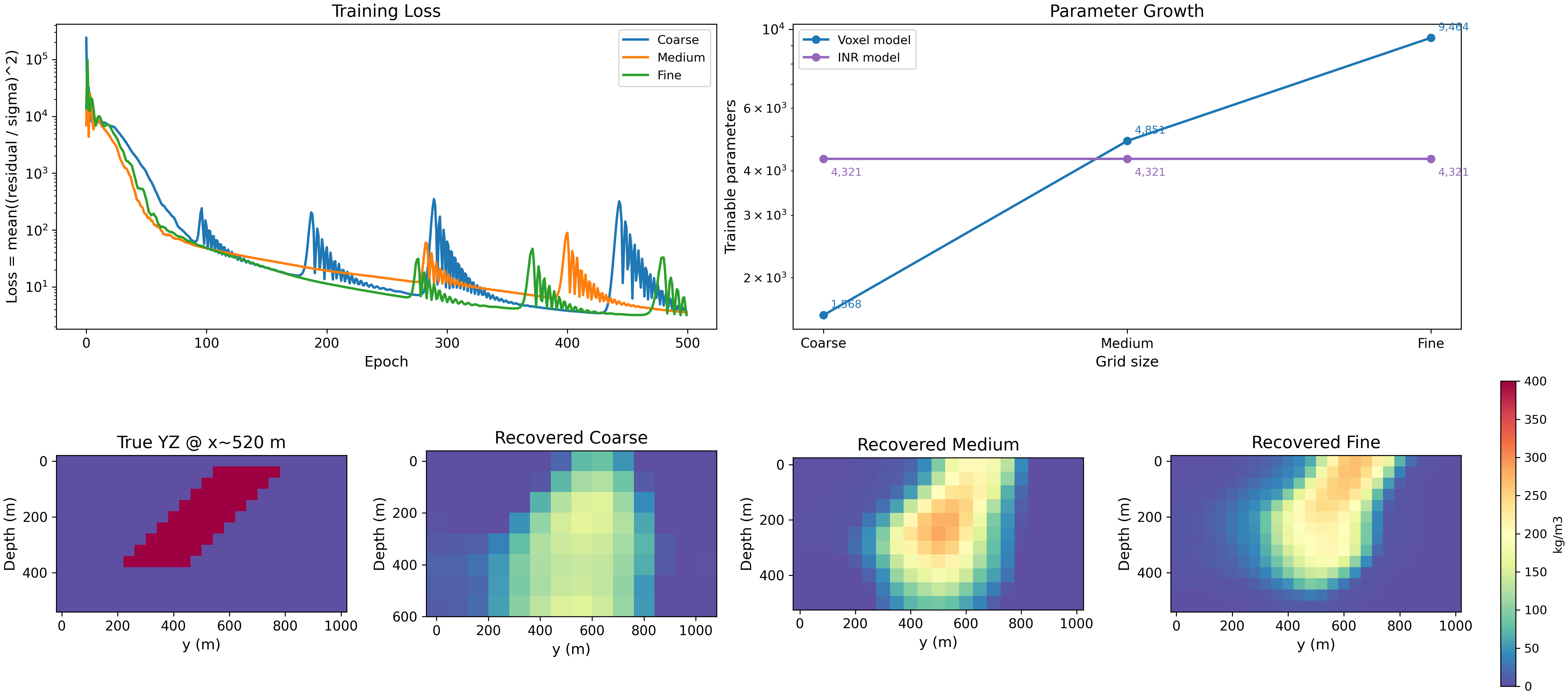}
    \caption{Effect of grid refinement on 3D gravity inversion with a fixed INR architecture. Top row: training-loss histories and comparison of voxel-model and INR trainable parameter counts for coarse, medium, and fine discretisations of the same nominal block-model domain. Bottom row: true $y$--$z$ cross-section and recovered $y$--$z$ cross-sections for the three grid sizes. The INR architecture is kept fixed across all runs, so its parameter count remains constant, whereas the voxel-model parameter count increases with grid refinement.}
    \label{fig:BlockGridSizeComparison}
\end{figure}

The second test addresses the complementary question of how discretisation affects the inversion when the INR architecture is kept fixed. To examine this, we repeat the same nominal block-model experiment on coarse, medium, and fine grids while using the same positional encoding and the same hidden-layer configuration $[48,48,24]$ in all cases, corresponding to $4{,}321$ trainable INR parameters. The target geometry, encoding, loss function, optimiser, and early-stopping rule are unchanged; only the discretisation is refined. In the present implementation, grid refinement also changes the observation layout because the surface survey is generated from the same coordinate arrays as the model grid, but this does not affect the main parameter-scaling argument.

Figure~\ref{fig:BlockGridSizeComparison} highlights the resulting contrast. As the discretisation is refined, the voxel-model parameter count increases directly with the number of cells, whereas the INR parameter count remains fixed because the architecture is unchanged. Thus, refinement of the spatial grid does not force a corresponding increase in INR trainable parameters. This is the key structural difference from voxel-based inversion, in which each refinement directly increases the number of unknowns.

The lower row of Figure~\ref{fig:BlockGridSizeComparison} shows that the recovered $y$--$z$ sections still depend on discretisation, because the continuous target is sampled differently on each grid and the forward response is represented at different resolutions. The coarse case produces the most diffuse reconstruction, whereas the medium and fine cases recover the dipping body more clearly. This trend should, however, be interpreted cautiously: because finer grids also imply denser sampling of the forward problem, the observed changes cannot be attributed solely to the fixed INR parameter count. The principal conclusion of this test is therefore not that finer grids necessarily improve inversion quality, but that discretisation refinement and parameter growth are decoupled in the INR formulation.

These two experiments provide a clearer picture of how INR inversion differs from voxel inversion. The network-size comparison shows that, on a fixed grid, increasing INR capacity improves reconstruction quality and data fit without requiring voxel-like parameter growth. The grid-size comparison shows that, for a fixed INR, refining the discretisation changes the sampled representation of the target and the forward problem, but does not automatically increase the number of trainable INR parameters. In this sense, the two tests isolate two distinct controls on inversion behaviour: INR capacity and discretisation.

This distinction is important for geophysical inverse problems. In voxel-based inversion, refining the model grid and enlarging the model space are effectively the same operation, since each additional cell introduces another unknown. In INR inversion, these controls are separate: one may refine the discretisation without increasing the number of trainable parameters, or increase the network capacity without altering the underlying grid. The combined results of these two experiments therefore show that INRs offer a more flexible relationship between discretisation, model complexity, and parameter count than conventional voxel parameterisations.

\subsection*{Implicit versus explicit regularisation}

Classical deterministic gravity inversion is commonly formulated as a stabilised least-squares problem because the mapping from density contrast to gravity is ill-posed and depth-decaying. Without regularisation, solutions tend to concentrate near the surface and become unstable to noise. Quadratic smoothness regularisation (first-order finite-difference penalties) suppresses oscillations but blurs sharp interfaces, while total variation (TV) regularisation promotes compact, blocky structure at the expense of increased nonlinearity and algorithmic complexity. Depth weighting, typically \(w(z)=(z+z_0)^{-\beta}\), is used to counteract the decay of sensitivity with depth \parencite{li19983}. Increasing smoothness and smallness stabilises the inversion but spreads density and reduces amplitudes, whereas stronger depth weighting restores deeper structure at the risk of amplifying noise and requiring tuning of \(\beta\) and \(z_0\) \parencite{last1983compact,li19983}.

The purpose of this experiment is to compare implicit regularisation arising from an implicit neural representation (INR) with conventional explicit regularisation strategies under controlled conditions. The test is carried out on a synthetic 3D gravity inversion problem defined on a \(1000\times1000\times500~\mathrm{m}\) domain discretised into \(50~\mathrm{m}\) cells. The true density model is a staircase-shaped block with contrast \(400~\mathrm{kg\,m^{-3}}\) embedded in a zero-background medium. Gravity data are generated at the surface on a regular grid using the same forward operator for all methods, and Gaussian noise with standard deviation \(0.01\,\mathrm{std}(\mathbf{d}_{\mathrm{true}})\) is added. All inversions are evaluated in whitened units, such that the target misfit corresponds to \(\mathrm{mean}[(r/\sigma)^2]\approx 1\).

Three inversion strategies are compared. The first uses an implicit neural representation with sinusoidal positional encoding (two frequency bands) and a compact multilayer perceptron with hidden layers \([48,48,24]\). The network output is bounded to \(\pm 600~\mathrm{kg\,m^{-3}}\). Training is performed using Adam with learning rate \(10^{-2}\) for up to 500 epochs. The loss consists only of the whitened data misfit, and early stopping is applied using a discrepancy-principle criterion: training is terminated once the weighted misfit reaches a tolerance band around unity. No explicit smoothness, smallness, or depth weighting is included. In this formulation, regularisation is controlled implicitly by the network parametrisation and the stopping rule.

The second inversion uses quadratic smoothness regularisation \parencite{menke2018geophysical}. It is formulated as a voxel-based Tikhonov regularised least-squares problem combining the data misfit with a depth-weighted smallness term and first-order finite-difference penalties in all spatial directions. The resulting linear system is solved iteratively using a conjugate-gradient (CG) method. The regularisation strength is tuned automatically so that the final weighted misfit matches the same noise-consistent target used for the INR.

The third inversion uses total variation (TV) regularisation \parencite{portniaguine1999focusing}. Because the TV penalty is non-quadratic, it is implemented using an iteratively reweighted least-squares (IRLS) scheme. At each IRLS iteration, a weighted least-squares system is constructed and solved using the same conjugate-gradient method. As in the smoothness case, the regularisation strength is adjusted so that the final misfit is consistent with the noise level.

\begin{figure}[H]
    \centering
    \includegraphics[width=\linewidth]{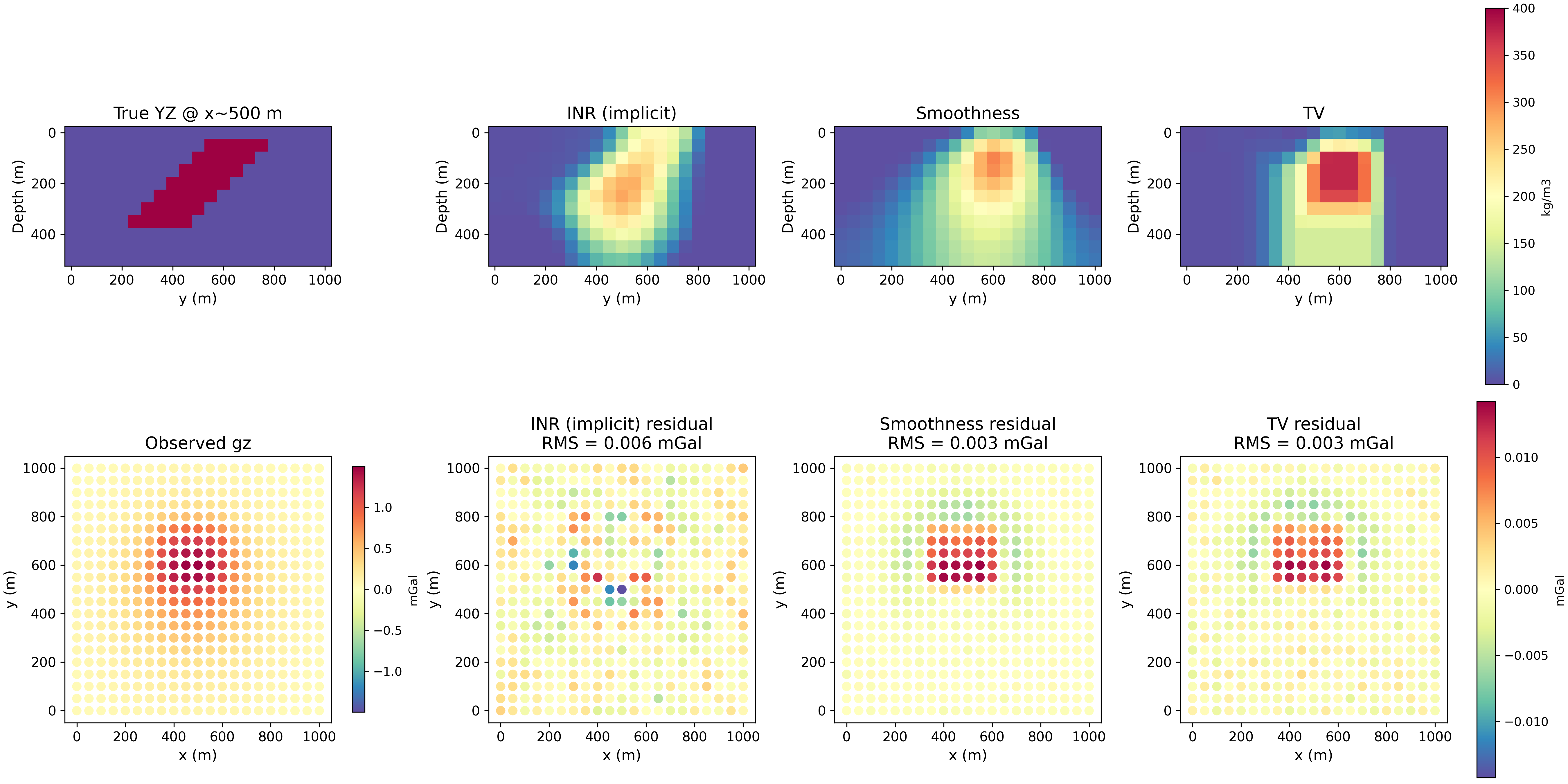}
    \caption{Comparison of implicit and explicit regularisation strategies for 3D gravity inversion. Top row: true density slice and recovered \(y\)–\(z\) slices from INR (implicit), smoothness regularisation, and total variation (TV). Bottom row: observed gravity data and corresponding residual maps. All methods are compared at similar noise-consistent data-fit levels.}
    \label{fig:ImplicitVsExplicit}
\end{figure}

Figure~\ref{fig:ImplicitVsExplicit} compares a representative \(y\)–\(z\) slice through the recovered models together with the observed data and residuals. All three methods achieve comparable data fits, with weighted misfit values close to unity, indicating that differences in the recovered models arise primarily from the regularisation strategy rather than unequal levels of data fit.

The INR reconstruction recovers the main geometry of the staircase block, including its lateral extent and depth variation, while remaining stable without any explicit penalty term. The recovered structure is smooth at the cell scale but preserves the overall blocky morphology and dip. Peak contrast is slightly underestimated and edges are rounded over a few cells, consistent with finite network capacity and the absence of an explicit focusing term.

The smoothness-regularised inversion produces a diffuse anomaly with rounded boundaries and reduced amplitude. The quadratic penalty spreads density beyond the true block extent and introduces vertical smearing despite the use of depth weighting \parencite{li19983,last1983compact}. The TV inversion yields a more compact structure with sharper contrasts than smoothness regularisation, but still simplifies the stepped geometry and introduces mild artefacts associated with the piecewise-constant prior.

These results show that the INR parametrisation acts as an implicit regulariser: the combination of limited network capacity, positional encoding, and early stopping favours spatially coherent solutions that fit the data without requiring explicit penalty terms. In contrast, the voxel-based approaches require explicit regularisation design and tuning of hyperparameters to achieve comparable stability.

The comparison is carried out at a fixed data-fit level rather than fixed regularisation strength. The stopping criterion in the INR plays a role analogous to regularisation-parameter selection in deterministic inversion, while the smoothness and TV methods adjust their regularisation weights to match the same target misfit. The experiment is limited to a single synthetic geometry and noise level, and gravity data remain band-limited, so sharp interfaces are not uniquely constrained. The results therefore demonstrate the effectiveness of implicit regularisation in this setting, but do not remove the need for careful control of model complexity in more general applications.

The results of this numerical test confirm the implicit regularisation capabilities of neural field parameterisation, observed in a recent synthetic study by \textcite{xu2025towards} for DC resistivity and seismic inversion experiments albeit in a 2D setting.  It should be noted that the baseline shown here is included solely to illustrate the implicit regularisation effect of the INR inversion. It is not intended as a quantitative benchmark against state-of-the-art deterministic methods; single-case results can overstate performance and are not necessarily generalisable.

\subsection*{Model variability and ensemble analysis}
The INR formulation introduces stochastic elements through random network initialisation and optimisation dynamics. To assess the stability of the recovered models under these factors, we perform an ensemble analysis in which multiple INR inversions are carried out for the same physical problem while varying only the initial network weights.
The experiment uses the same synthetic block-model setting as the preceding examples. The domain is \(1000\times1000\times500~\mathrm{m}\), discretised with \(50~\mathrm{m}\) cells. The true density consists of a staircase-shaped body with contrast \(400~\mathrm{kg\,m^{-3}}\) embedded in a zero-background medium. Gravity data are generated at the surface using the same forward operator and contaminated with Gaussian noise of standard deviation \(0.01\,\mathrm{std}(\mathbf{d}_{\mathrm{true}})\). The observed data are held fixed across all runs.
Each ensemble member uses the same INR configuration: sinusoidal positional encoding with two frequency bands and a four-layer multilayer perceptron with hidden widths \([256,256,256,256]\). Training is performed using Adam with learning rate \(10^{-2}\) for up to 500 epochs, and early stopping is applied using the same discrepancy-principle criterion as in previous experiments, targeting \(\mathrm{mean}[(r/\sigma)^2]\approx 1\). The only difference between ensemble members is the random initialisation of the network parameters. This design isolates variability arising from optimisation and parametrisation rather than changes in data or forward modelling.

\begin{figure}[H]
    \centering
    \includegraphics[width=\linewidth]{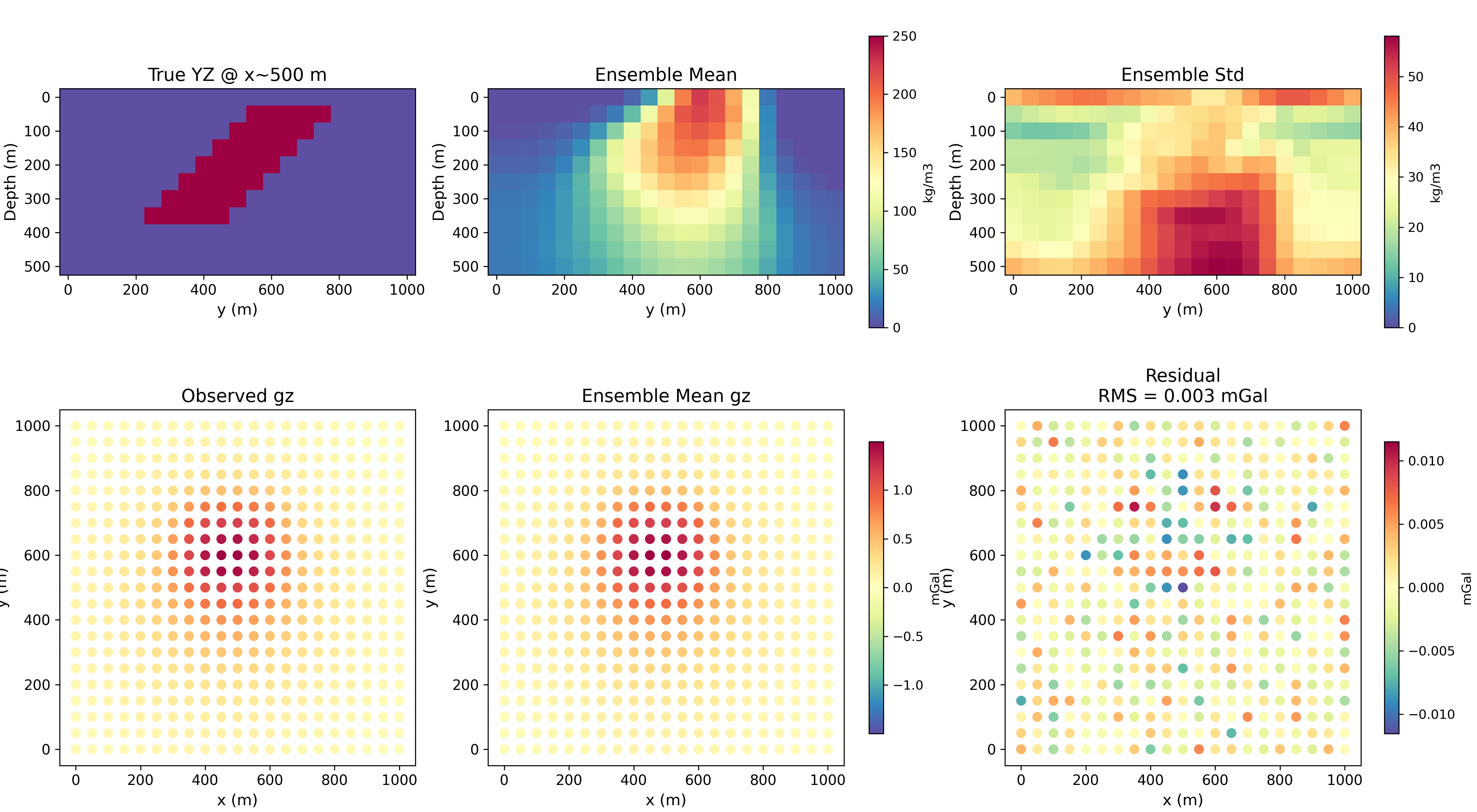}
    \caption{Ensemble statistics for INR inversion. Top row: true model, ensemble mean, and ensemble standard deviation for a representative \(y\)–\(z\) slice. Bottom row: observed gravity data, predicted gravity from the ensemble mean model, and residuals.}
    \label{fig:ModelEnsembleMean}
\end{figure}
Figure~\ref{fig:ModelEnsembleMean} shows the ensemble mean and standard deviation of the recovered density models. The ensemble mean recovers the main geometry of the block, including its lateral extent and depth variation, and produces a data fit consistent with the noise level. The ensemble standard deviation highlights regions of higher uncertainty, which are concentrated around the boundaries of the block and at greater depths where sensitivity is weaker. In contrast, the interior of the block and the background regions show relatively low variability, indicating that these features are consistently recovered across different initialisations.
\begin{figure}[H]
    \centering
    \includegraphics[width=0.8\linewidth]{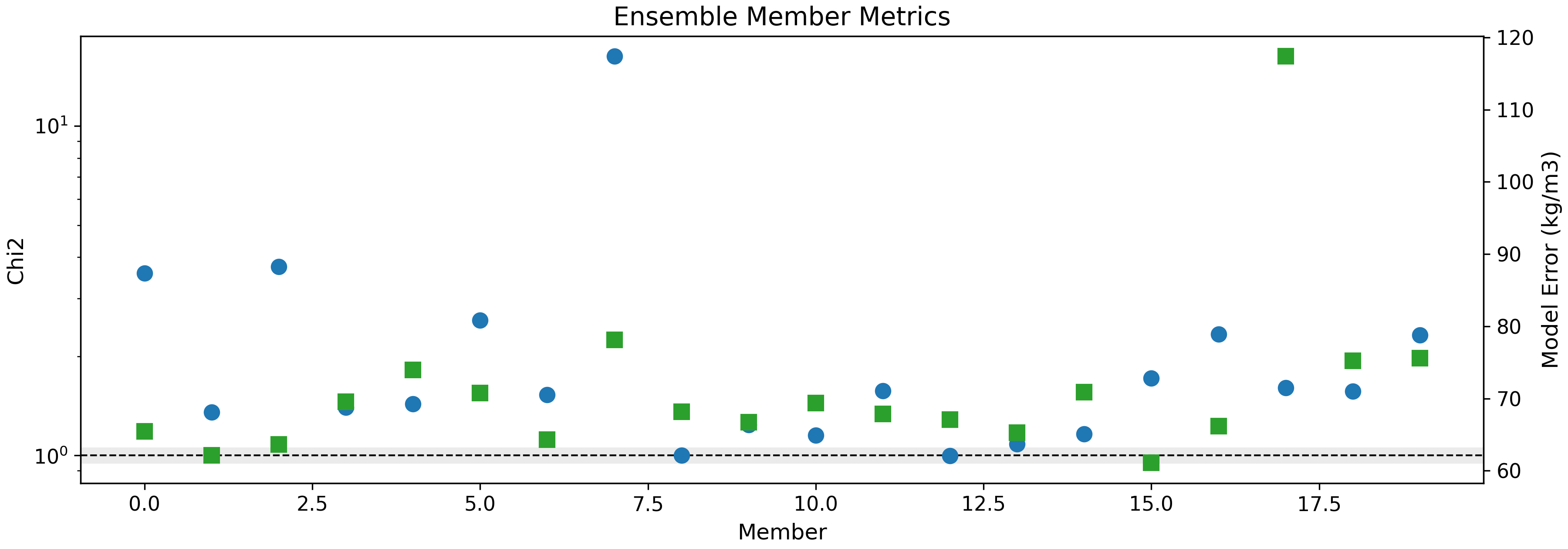}
    \caption{Per-member ensemble metrics. Blue circles show the weighted data misfit (\(\chi^2\)), and green squares show the RMS model error relative to the true density. The dashed line indicates the target misfit level.}
    \label{fig:ModelEnsembleMetrics}
\end{figure}

Figure~\ref{fig:ModelEnsembleMetrics} summarises per-member performance. All ensemble members converge to weighted misfit values close to unity, consistent with the early stopping criterion, indicating stable data fitting across different initialisations. However, the model-space error varies between members, demonstrating that multiple density distributions can explain the data equally well. This behaviour reflects the non-uniqueness of the gravity inversion problem and the implicit prior imposed by the INR parametrisation.
The ensemble results show that INR inversion produces consistent large-scale structures while allowing variability in regions that are weakly constrained by the data. The spread of the ensemble provides a practical measure of uncertainty that arises from the inversion process itself. At the same time, the variability between models, despite similar data misfits, highlights that the recovered solution is not unique and depends on the implicit biases of the network and optimisation. This ensemble approach therefore provides a useful framework for assessing the reliability and stability of INR-based gravity inversion results.

\subsection*{Limitations and potential sources of bias}
\begin{enumerate}
    \item[1] The present study is based on controlled synthetic experiments, which introduce several limitations that should be considered when interpreting the results. First, the use of synthetic density models (Gaussian random fields and block geometries) does not fully capture the structural complexity, noise characteristics, and prior uncertainty present in field data. As a result, the reported performance may be optimistic compared to real-world applications.     
    \item[2] The noise model is prescribed and relatively simple. Although Gaussian, correlated, and sparse outlier noise are considered, all cases assume stationary statistics and known noise levels (See supplementary material). In practice, gravity data often exhibit spatially varying and poorly characterised errors, which may affect inversion stability and parameter selection.
    \item[3] The INR inversion relies on implicit regularisation controlled by the neural-network parametrisation, spatial encoding, and early stopping. These choices introduce an implicit prior that is not explicitly defined and may bias the solution toward smooth or spatially coherent structures. 
    \item[4] Although, INR method potentially omits the traditional regularisation and depth weighting, it introduced new hyperparameters related to network architecture that need to be selected manually.  
    \item[5] The forward model links the continuous INR representation to a discretised prism-based operator evaluated at cell centres. This introduces a discretisation bias, as within-cell density variations are not represented in the physics. Consequently, the recovered model depends on the chosen grid resolution and may not fully reflect sub-cell structure. 
    \item[6] The experimental scope is limited to a small set of synthetic scenarios and noise levels. While these tests illustrate key behaviours of the method, they do not provide a comprehensive benchmark against all existing inversion strategies. Further validation on field datasets and across a wider range of geological settings is required to assess general applicability. The experiments in this paper are designed to isolate INR method behaviour rather than provide exhaustive benchmarking.
\end{enumerate}

\section*{Conclusions}
This study shows that implicit neural representations (INRs) provide an effective continuous parameterisation for three-dimensional gravity inversion. With positional encoding, the network represents both long and short wavelengths and mitigates the spectral bias of plain coordinate MLPs that otherwise smear sharp boundaries. Network capacity acts as an implicit regulariser: shared weights, limited degrees of freedom, smooth activations and early stopping favour parsimonious, coherent structure unless the data require additional detail. Voxel inversions allocate many independent parameters near the surface, and the decay of gravitational sensitivity then biases structure upward unless a depth-weighting law (e.g., parameters $\beta$, $z_0$) is tuned. The INR does not assign cell-wise freedom; its basis spans the volume uniformly. To match the observations, the model places contrast where the forward physics demand it, including at depth, without user-imposed weighting. Presented numerical experiments support these observations. Positional encoding improves recovery of high-frequency geology; increasing capacity reduces misfit with diminishing returns near the noise floor. In a block test, the INR recovers sharp lateral boundaries and credible depth extent using only a data-misfit objective.

The scope of this paper is intentionally restricted to controlled synthetics with known physics and noise. The results therefore open several questions that require further testing, including application to a well-studied field dataset with independent geological and geophysical constraints. Non-uniqueness persists, and the practical performance of the approach will depend on choices that must respect station spacing, depth sensitivity and noise level. Although the framework omits explicit tuning of explicit regularisation and depth-weighting parameters, it introduces a new set of hyperparameters that shape the implicit prior: the bandwidth and number of positional-encoding frequencies; network capacity (depth and width), activation functions and output bounds; optimiser and schedule (learning rate, batch size, early stopping); normalisation of coordinates and data; random initialisation; and the forward-model discretisation that links the continuous INR to the prism physics. These settings govern the balance between stability and detail: excessive bandwidth can introduce weakly constrained texture; insufficient capacity can over-smooth resolvable structure; coarse forward grids can mask within-cell variability. Future work will broaden quantitative comparisons with established voxel and unstructured-mesh inversions, develop principled guidance for bandwidth and capacity selection (e.g., tied to survey Nyquist limits and depth kernels), assess robustness across survey layouts and noise models, and add uncertainty characterisation using ensembles or approximate Bayesian methods. Even with these limitations, INRs emerge here as promising, flexible and geologically sensible model representations for large-scale gravity inversion, and they offer a clear path toward diverse applications in geophysical inversion.

\section*{Code Availability Statement}
The codes for reproducing the results can be found at \href{https://zenodo.org/records/19440024}{https://zenodo.org/records/19440024} 

\section*{Author contributions}
Conceptualisation: PKM; Code development: PKM; Numerical experiments: PKM and SL; Manuscript writing/editing: PKM,  JK, AS, and SL; Funding acquisition: JK, PKM. All authors reviewed the manuscript.

\section*{Acknowledgment}
This work was supported by the Research Council of Finland (project 359261). The authors wish to acknowledge CSC – IT Center for Science, Finland, for computational resources. 

\section*{Declaration of AI usage}
GPT-5.4 (OpenAI) was used in a limited role during manuscript polishing and code cleanup and structuring, primarily for language editing, formatting, and minor text refinement. All methodological decisions, analyses, implementations, and conclusions were carried out and verified by the authors.

\renewcommand\refname{References}
\setlength{\bibitemsep}{0pt}
\renewcommand{\bibfont}{\footnotesize}
\printbibliography

@article{li19983,
  title   = {3-D inversion of gravity data},
  author  = {Li, Yaoguo and Oldenburg, Douglas W.},
  journal = {Geophysics},
  volume  = {63},
  number  = {1},
  pages   = {109--119},
  year    = {1998}
}

@article{oldenburg1991inversion,
  title   = {Inversion of geophysical data using an approximate inverse mapping},
  author  = {Oldenburg, Douglas W. and Ellis, Richard G.},
  journal = {Geophysical Journal International},
  volume  = {105},
  number  = {2},
  pages   = {325--353},
  year    = {1991}
}

@article{last1983compact,
  title   = {Compact gravity inversion},
  author  = {Last, B. J. and Kubik, K.},
  journal = {Geophysics},
  volume  = {48},
  number  = {6},
  pages   = {713--721},
  year    = {1983}
}

@article{parker1975theory,
  title   = {The theory of ideal bodies for gravity interpretation},
  author  = {Parker, Robert L.},
  journal = {Geophysical Journal International},
  volume  = {42},
  number  = {2},
  pages   = {315--334},
  year    = {1975}
}

@article{huang2021deep,
  title={Deep learning 3D sparse inversion of gravity data},
  author={Huang, Rui and Liu, Shuang and Qi, Rui and Zhang, Yujie},
  journal={Journal of Geophysical Research: Solid Earth},
  volume={126},
  number={11},
  pages={e2021JB022476},
  year={2021},
  publisher={Wiley Online Library}
}

@article{yang20213,
  title={3-D gravity inversion based on deep convolution neural networks},
  author={Yang, Qianguo and Hu, Xiangyun and Liu, Shuang and Jie, Qu and Wang, Huaijiang and Chen, Qiuhua},
  journal={IEEE geoscience and remote sensing letters},
  volume={19},
  pages={1--5},
  year={2021},
  publisher={IEEE}
}

@article{yang2022deep,
  title={Deep learning inversion of gravity data for detection of CO2 plumes in overlying aquifers},
  author={Yang, Xianjin and Chen, Xiao and Smith, Megan M},
  journal={Journal of Applied Geophysics},
  volume={196},
  pages={104507},
  year={2022},
  publisher={Elsevier}
}

@article{wu2023improved,
  title={Improved gravity inversion method based on deep learning with physical constraint and its application to the airborne gravity data in East Antarctica},
  author={Wu, Guochao and Wei, Yue and Dong, Siyuan and Zhang, Tao and Yang, Chunguo and Qin, Linjiang and Guan, Qingsheng},
  journal={Remote Sensing},
  volume={15},
  number={20},
  pages={4933},
  year={2023},
  publisher={MDPI}
}

@ARTICLE{9965416,
  author={Li, Yinshuo and Jia, Zhuo and Lu, Wenkai},
  journal={IEEE Transactions on Geoscience and Remote Sensing}, 
  title={Self-Supervised Deep Learning for 3D Gravity Inversion}, 
  year={2022},
  volume={60},
  number={},
  pages={1-11}
}

@article{cai2025effective,
  title={Effective gravity inversion of basement relief with unfixed density contrast using deep learning},
  author={Cai, Hongzhu and He, Siyuan and He, Ziang and Liu, Shuang and Liu, Lichao and Hu, Xiangyun},
  journal={Computers \& Geosciences},
  volume={196},
  pages={105832},
  year={2025},
  publisher={Elsevier}
}

@article{zhang2022decnet,
  title={DecNet: Decomposition network for 3D gravity inversion},
  author={Zhang, Shuang and Yin, Changchun and Cao, Xiaoyue and Sun, Siyuan and Liu, Yunhe and Ren, Xiuyan},
  journal={Geophysics},
  volume={87},
  number={5},
  pages={G103--G114},
  year={2022},
  publisher={Society of Exploration Geophysicists}
}

@ARTICLE{10034785,
  author={Zhou, Xinyi and Chen, Zhaoxi and Lv, Yandong and Wang, Shuai},
  journal={IEEE Transactions on Geoscience and Remote Sensing}, 
  title={3-D Gravity Intelligent Inversion by U-Net Network With Data Augmentation}, 
  year={2023},
  volume={61},
  number={},
  pages={1-13},
  doi={10.1109/TGRS.2023.3241310}
}

@article{yu2021three,
  title={Three-dimensional gravity inversion based on 3D U-Net++},
  author={Yu-Feng, Wang and Yu-Jie, Zhang and Li-Hua, Fu and Hong-Wei, Li},
  journal={Applied Geophysics},
  volume={18},
  number={4},
  pages={451--460},
  year={2021},
  publisher={Springer}
}

@article{sitzmann2020implicit,
  title={Implicit neural representations with periodic activation functions},
  author={Sitzmann, Vincent and Martel, Julien and Bergman, Alexander and Lindell, David and Wetzstein, Gordon},
  journal={Advances in neural information processing systems},
  volume={33},
  pages={7462--7473},
  year={2020}
}

@article{tancik2020fourier,
  title={Fourier features let networks learn high frequency functions in low dimensional domains},
  author={Tancik, Matthew and Srinivasan, Pratul and Mildenhall, Ben and Fridovich-Keil, Sara and Raghavan, Nithin and Singhal, Utkarsh and Ramamoorthi, Ravi and Barron, Jonathan and Ng, Ren},
  journal={Advances in neural information processing systems},
  volume={33},
  pages={7537--7547},
  year={2020}
}

@article{muller2022instant,
  title={Instant neural graphics primitives with a multiresolution hash encoding},
  author={M{\"u}ller, Thomas and Evans, Alex and Schied, Christoph and Keller, Alexander},
  journal={ACM transactions on graphics (TOG)},
  volume={41},
  number={4},
  pages={1--15},
  year={2022},
  publisher={ACM New York, NY, USA}
}

@inproceedings{rahaman2019spectral,
  title={On the spectral bias of neural networks},
  author={Rahaman, Nasim and Baratin, Aristide and Arpit, Devansh and Draxler, Felix and Lin, Min and Hamprecht, Fred and Bengio, Yoshua and Courville, Aaron},
  booktitle={International conference on machine learning},
  pages={5301--5310},
  year={2019},
  organization={PMLR}
}

@article{nagy2000gravitational,
  title={The gravitational potential and its derivatives for the prism},
  author={Nagy, Dezsö and Papp, Gabor and Benedek, Judit},
  journal={Journal of Geodesy},
  volume={74},
  number={7},
  pages={552--560},
  year={2000},
  publisher={Springer}
}

@article{Nagy1966,
author = {Dezsö Nagy},
title = {The gravitational attraction of a right rectangular prism},
journal = {GEOPHYSICS},
volume = {31},
number = {2},
pages = {362-371},
year = {1966},
doi = {10.1190/1.1439779}
}

@article{Faroughi2024,
    author = {Faroughi, Salah A. and Pawar, Nikhil M. and Fernandes, Célio and Raissi, Maziar and Das, Subasish and Kalantari, Nima K. and Kourosh Mahjour, Seyed},
    title = {Physics-Guided, Physics-Informed, and Physics-Encoded Neural Networks and Operators in Scientific Computing: Fluid and Solid Mechanics},
    journal = {Journal of Computing and Information Science in Engineering},
    volume = {24},
    number = {4},
    pages = {040802},
    year = {2024},
    month = {01}
}

@article{xu2025towards,
  title={Towards Understanding the Benefits of Neural Network Parameterizations in Geophysical Inversions: A Study With Neural Fields},
  author={Xu, Anran and Heagy, Lindsey J},
  journal={arXiv preprint arXiv:2503.17503},
  year={2025}
}

@article{sun2023implicit,
  title={Implicit seismic full waveform inversion with deep neural representation},
  author={Sun, Jian and Innanen, Kristopher and Zhang, Tianze and Trad, Daniel},
  journal={Journal of Geophysical Research: Solid Earth},
  volume={128},
  number={3},
  pages={e2022JB025964},
  year={2023},
  publisher={Wiley Online Library}
}

@misc{essakine2025,
      title={Where Do We Stand with Implicit Neural Representations? A Technical and Performance Survey}, 
      author={Amer Essakine and Yanqi Cheng and Chun-Wun Cheng and Lipei Zhang and Zhongying Deng and Lei Zhu and Carola-Bibiane Schönlieb and Angelica I Aviles-Rivero},
      year={2025},
      eprint={2411.03688},
      archivePrefix={arXiv},
      primaryClass={cs.CV},
      url={https://arxiv.org/abs/2411.03688}, 
}

@article{romero2025bayesian,
  title={Bayesian Seismic Inversion with Implicit Neural Representations},
  author={Romero, Juan and Heidrich, Wolfgang and Ravasi, Matteo},
  journal={Geophysical Journal International},
  pages={ggaf249},
  year={2025},
  publisher={Oxford University Press}
}

@article{smith2025implicit,
  title={Implicit neural representation for potential field geophysics},
  author={Smith, Luke Thomas and Horrocks, Tom and Akhtar, Naveed and Holden, Eun-Jung and Wedge, Daniel},
  journal={Scientific Reports},
  volume={15},
  number={1},
  pages={9799},
  year={2025},
  publisher={Nature Publishing Group UK London}
}

@article{Mishra2025,
    author = {Mishra, Pankaj K},
    title = {Lithosphere in digital twins of the Earth},
    journal = {Geophysical Journal International},
    volume = {242},
    number = {2},
    pages = {ggaf199},
    year = {2025},
    month = {05}
}

@article{danaei20223d,
  title={3D inversion of gravity data with unstructured mesh and least-squares QR-factorization (LSQR)},
  author={Danaei, Khatereh and Moradzadeh, Ali and Norouzi, Gholam-Hossain and Smith, Richard and Abedi, Maysam and Fam, Hossein Jodeiri Akbari},
  journal={Journal of Applied Geophysics},
  volume={206},
  pages={104781},
  year={2022},
  publisher={Elsevier}
}

@article{davis2013,
  title={Efficient 3D inversion of magnetic data via octree-mesh discretization, space-filling curves, and wavelets},
  author={Davis, Kristofer and Li, Yaoguo},
  journal={Geophysics},
  volume={78},
  number={5},
  pages={J61--J73},
  year={2013},
  publisher={Society of Exploration Geophysicists}
}

@article{liu2019advances,
  title={Advances in Gaussian random field generation: a review},
  author={Liu, Yang and Li, Jingfa and Sun, Shuyu and Yu, Bo},
  journal={Computational Geosciences},
  volume={23},
  number={5},
  pages={1011--1047},
  year={2019},
  publisher={Springer}
}

@article{mishra2025stochastic,
  title={Stochastic Joint Inversion of Seismic and Controlled-Source Electromagnetic Data},
  author={Mishra, Pankaj K and Arnulf, Adrien and Sen, Mrinal K and Zhao, Zeyu and Jaysaval, Piyoosh},
  journal={Geophysical Prospecting},
  volume={73},
  number={6},
  pages={e70043},
  year={2025},
  publisher={European Association of Geoscientists \& Engineers}
}

@article{mildenhall2021nerf,
  title={Nerf: Representing scenes as neural radiance fields for view synthesis},
  author={Mildenhall, Ben and Srinivasan, Pratul P and Tancik, Matthew and Barron, Jonathan T and Ramamoorthi, Ravi and Ng, Ren},
  journal={Communications of the ACM},
  volume={65},
  number={1},
  pages={99--106},
  year={2021},
  publisher={ACM New York, NY, USA}
}

@article{izzo2022geodesy,
  title={Geodesy of irregular small bodies via neural density fields},
  author={Izzo, Dario and G{\'o}mez, Pablo},
  journal={Communications Engineering},
  volume={1},
  number={1},
  pages={48},
  year={2022},
  publisher={Nature Publishing Group UK London}
}

@article{schuhmacher2023investigation,
  title={Investigation of the robustness of neural density fields},
  author={Schuhmacher, Jonas and Gratl, Fabio and Izzo, Dario and G{\'o}mez, Pablo},
  journal={arXiv preprint arXiv:2305.19698},
  year={2023}
}

@article{li2025implicit,
  title={Implicit neural representations for 3D gravity inversion},
  author={Li, Xiong and Zhao, Jingtao and Zhou, Shuai},
  journal={Computers \& Geosciences},
  pages={106082},
  year={2025},
  publisher={Elsevier}
}

@article{hillier2023geoinr,
  title={GeoINR 1.0: an implicit neural network approach to three-dimensional geological modelling},
  author={Hillier, Michael and Wellmann, Florian and de Kemp, Eric A and Brodaric, Boyan and Schetselaar, Ernst and Bedard, Karine},
  journal={Geoscientific Model Development},
  volume={16},
  number={23},
  pages={6987--7012},
  year={2023},
  publisher={Copernicus GmbH}
}

@inproceedings{xu2020reluplex,
  title={Reluplex made more practical: Leaky ReLU},
  author={Xu, Jin and Li, Zishan and Du, Bowen and Zhang, Miaomiao and Liu, Jing},
  booktitle={2020 IEEE Symposium on Computers and communications (ISCC)},
  pages={1--7},
  year={2020},
  organization={IEEE}
}

@article{kingma2014adam,
  title={Adam: A method for stochastic optimization},
  author={Kingma, Diederik P and Ba, Jimmy},
  journal={arXiv preprint arXiv:1412.6980},
  year={2014}
}

@article{portniaguine1999focusing,
  title={Focusing geophysical inversion images},
  author={Portniaguine, Oleg and Zhdanov, Michael S},
  journal={Geophysics},
  volume={64},
  number={3},
  pages={874--887},
  year={1999},
  publisher={Society of Exploration Geophysicists}
}

@book{menke2018geophysical,
  title={Geophysical data analysis: Discrete inverse theory},
  author={Menke, William},
  year={2018},
  publisher={Academic press}
}

\label{lastpage}

\end{document}


\linenumbers
\begin{center}
 \textbf{\Huge{Supplementary material}}   
\end{center}

\textbf{\huge{Three-dimensional inversion of gravity data using implicit neural representations and scientific machine learning}}

\vspace{0.5cm}
\large \textbf{Pankaj K Mishra} \orcidlink{0000-0003-4907-4724}$^{1\ast}$, \textbf{Sanni Laaksonen}$^{1}$, \textbf{Jochen Kamm}$^{1}$, \textbf{Anand Singh}$^{2}$

\vspace{0.2cm} 
\small
$^{1}$Geological Survey of Finland, Vuorimiehentie 5, Espoo, Finland\\
$^{2}$Indian Institute of Technology Bombay, Mumbai, India\\
$^{\ast}$Corresponding author: \href{mailto:pankaj.mishra@gtk.fi}{pankaj.mishra@gtk.fi}
\noindent\rule{\textwidth}{0.2pt}
\section*{Noise Sensitivity of the INR Inversion}

To evaluate the robustness of the INR-based gravity inversion to data contamination, we carried out a controlled noise-sensitivity study using the same synthetic block-model setup as in the previous experiments. The computational domain spans $1000 \times 1000 \times 500$~m and is discretized with 50~m cells. The true density model consists of a staircase-shaped body with density contrast $400~\mathrm{kg/m^3}$ embedded in a zero-background medium. Gravity data are computed on a regular surface grid using the same forward operator as in the preceding tests. 

To isolate the effect of noise alone, the inversion configuration was kept fixed across all experiments: the INR architecture, positional encoding (two frequency bands), optimiser (Adam), learning rate, training schedule, and early-stopping criterion were identical. Only the noise model and noise amplitude were varied. The stopping rule follows a discrepancy principle, targeting a whitened data misfit $\chi^2 = \mathrm{mean}[(r/\sigma)^2] \approx 1$.

Three noise classes were considered. The first is Gaussian noise, which serves as the reference case. The second is spatially correlated noise, introduced to mimic survey and processing effects such as drift, leveling errors, or imperfect background removal. The third is sparse outlier noise, representing localized corrupted observations. Each noise type was tested at amplitudes of 2\%, 5\%, and 10\% of the standard deviation of the noise-free gravity response, extending the 1\% baseline used in earlier experiments.

\begin{figure*}[t]
    \centering
    \includegraphics[width=\textwidth]{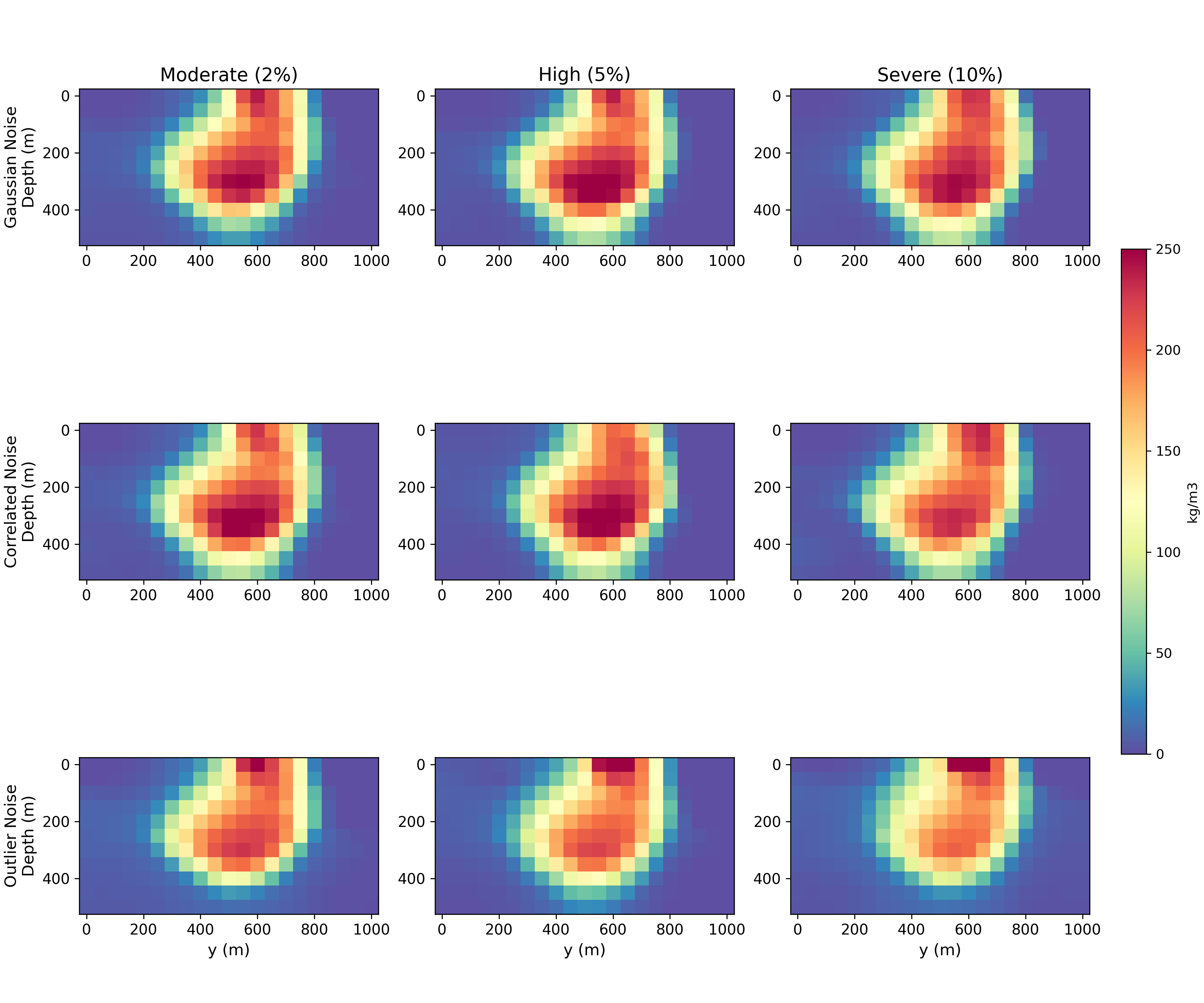}
    \caption{
    Recovered $y$–$z$ sections from the INR inversion for three noise classes (Gaussian, correlated, and outlier noise) and three noise levels (2\%, 5\%, and 10\%). The main geometry and location of the staircase-shaped target are preserved across all tested conditions. Gaussian and correlated noise yield very similar reconstructions, whereas outlier contamination produces somewhat stronger degradation in amplitude and boundary sharpness.
    }
    \label{fig:noise_sensitivity_models}
\end{figure*}

Figure~\ref{fig:noise_sensitivity_models} summarises the recovered models. The $y$–$z$ sections show that the INR inversion preserves the main geometry and location of the target body across all nine noise conditions. Variations appear primarily in amplitude and boundary sharpness as noise increases. The Gaussian and correlated noise cases produce very similar reconstructions, indicating that moderate spatial correlation in the data errors does not significantly affect the recovered structure. The outlier-noise cases show more visible degradation, but the anomaly remains clearly identifiable.

\begin{figure*}[t]
    \centering
    \includegraphics[width=\textwidth]{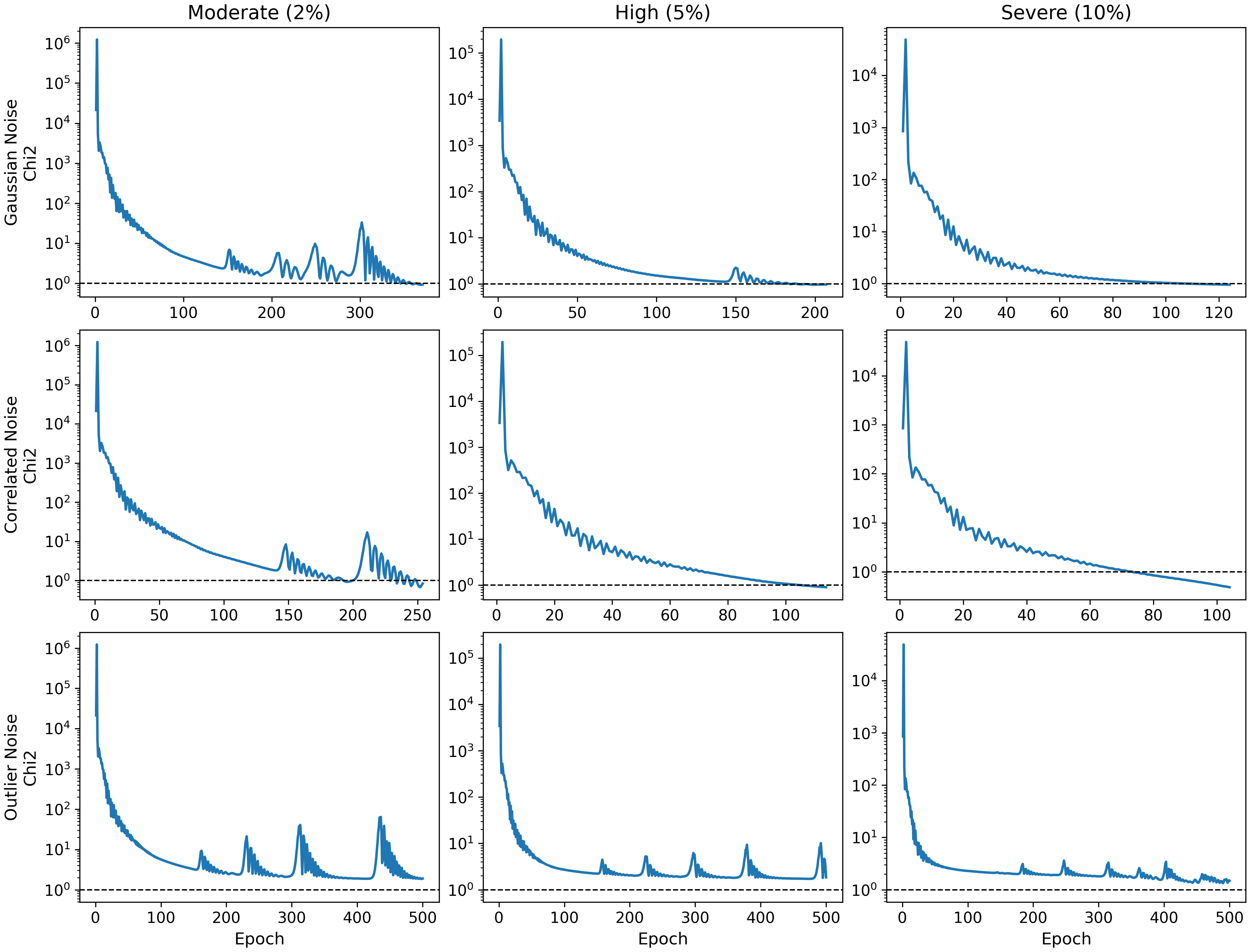}
    \caption{
    Convergence histories for the noise-sensitivity study, shown in terms of the weighted data misfit, $\chi^2$, for the same nine cases as in Figure~\ref{fig:noise_sensitivity_models}. The dashed horizontal line marks the target noise level. Gaussian and correlated noise cases converge stably toward the target, while outlier-noise cases show slower and more oscillatory behaviour, consistent with the sensitivity of the squared-error loss to isolated bad observations.
    }
    \label{fig:noise_sensitivity_convergence}
\end{figure*}

The convergence histories in Figure~\ref{fig:noise_sensitivity_convergence} provide additional insight into optimisation behaviour. For Gaussian and correlated noise, the weighted misfit $\chi^2$ decreases rapidly and approaches the target noise level in a stable manner. In contrast, the outlier-noise cases exhibit slower convergence and oscillatory behaviour. This is consistent with the use of a squared-error loss, which is known to be sensitive to outliers. Despite this, the inversion remains stable and converges to models that reproduce the dominant structure of the anomaly.

Quantitative metrics are consistent with the visual interpretation. Across the Gaussian and correlated cases, the final model errors vary only moderately with increasing noise amplitude. Even at the 10\% level, the recovered structures remain comparable in geometry to those obtained at lower noise levels. The outlier cases show higher model error and misfit variability, but the degradation is gradual rather than abrupt.

These results indicate that the INR inversion is robust across a range of realistic noise conditions. Performance is strongest for Gaussian and correlated noise, while sparse outliers produce the largest degradation, as expected for an $L_2$ data-misfit formulation. The method degrades progressively rather than failing, and continues to recover the principal subsurface structure even under substantially noisier data. This behaviour suggests that the INR framework does not rely on a narrowly idealised noise model and remains applicable under conditions relevant to practical gravity inversion.

\renewcommand\refname{References}
\setlength{\bibitemsep}{0pt}
\renewcommand{\bibfont}{\footnotesize}
\printbibliography

\label{lastpage}